\documentclass{aa}  

\usepackage{graphicx}
\usepackage{txfonts}
\usepackage{xcolor}
\usepackage[colorlinks=true,
            linkcolor=blue,
            urlcolor=blue,
            citecolor=blue]{hyperref}

\newcommand{\hkpc}{h^{-1}{\rm kpc}}
\newcommand{\hmsun}{h^{-1}{\rm M}_\odot}
\newcommand{\hmpc}{h^{-1}{\rm Mpc}}

\newcommand{\zobov}{\textsc{zobov}}
\newcommand{\revolver}{\textsc{revolver}}
\newcommand{\revolverzobov}{\textsc{revolver}-\textsc{zobov}}
\newcommand{\revolvervoxel}{\textsc{revolver}-\textsc{voxel}}
\newcommand{\popcorn}{\textsc{popcorn}}
\newcommand{\sparkling}{\textsc{sparkling}}

\begin{document} 
  
   \title{The Galaxy Bias Profile of Cosmic Voids:}

    \subtitle{A Comparison of Void Finders}

   \author{Ignacio G. Alfaro$^{1,2,3}$ \thanks{E-mail:german.alfaro@unc.edu.ar}, Antonio D. Montero-Dorta$^{1}$, Jorge F. Bustillos$^{2,3}$, Dante J. Paz $^{2,3}$, Andr\'es N. Ruiz$^{2,3}$, Andrés Balaguera-Antolínez, Ravi K. Sheth$^{4,5}$, Facundo Rodriguez$^{2,3}$, Constanza A. Soto-Suárez$^{6,7}$}
   
   \authorrunning{I. G. Alfaro et al.}
   
   \institute{
              Departamento de Física, Universidad Técnica Federico Santa María, Avenida Vicuña Mackenna 3939, San Joaquín, Santiago, Chile. \and
              Instituto de Astronomía Teórica y Experimental, CONICET-UNC, Laprida 854, X5000BGR, C\'ordoba, Argentina. \and 
              Observatorio Astron\'omico de C\'ordoba, UNC, Laprida 854, X5000BGR, C\'ordoba, Argentina. \and
            Center for Particle Cosmology, University of Pennsylvania, Philadelphia, PA 19104, USA. \and The Abdus Salam International Center for Theoretical Physics, Strada Costiera 11, Trieste 34151, Italy. \and
            Departamento de Física, Universidad Técnica Federico Santa María, Casilla 110-V, Avda. España 1680, Valparaíso, Chile. \and 
            Instituto de Física, Pontificia Universidad Católica de Valparaíso, Casilla 4950, Valparaíso, Chile.              
              }

   \date{\today}

  \abstract
   {Cosmic voids, the largest underdense regions in the Universe, provide unique laboratories for studying galaxy formation in extreme low-density environments and constitute powerful probes of cosmology. Recent work has shown that individual galaxy bias ($b_i$), which quantifies how each galaxy traces the underlying dark matter field, exhibits a characteristic radial dependence within spherical voids, defining a void bias profile in which galaxies near void centers display systematically lower bias values. However, the extent to which this bias profile depends on the void identification methodology has so far remained unexplored.}
   {We investigate how the environmental modulation of individual galaxy bias depends on the adopted void-finding algorithm by comparing measurements across five distinct void definitions: spherical voids (\sparkling), watershed-based methods (\zobov\ and \revolver\ in two modes), and free-form integrated-density voids (\popcorn).}
   {We apply these complementary void-finding algorithms to the same galaxy sample drawn from the IllustrisTNG simulation (TNG300-1 at $z=0$) and compute individual galaxy bias profiles as a function of distance from void centers. We quantify the correlation between $b_i$ and the membership of the void catalogs and explore how this relationship varies with the integrated underdensity threshold for density-based methods.}
   {We find that the radial gradient of individual bias within voids, generally increasing from negative values at the void centers to higher values at the boundaries, is robust across most void definitions. However, density-threshold methods (\sparkling\ and \popcorn) preferentially select galaxies with $b_i<0$, while watershed methods without density constraints (\zobov) include substantial contamination from high-bias boundary galaxies. The correlation between negative bias and void membership is systematically strengthened as the integrated underdensity threshold becomes less restrictive, with \popcorn\ achieving the highest purity in isolating anti-biased populations.}
   {}
   \maketitle
  
%
%
\section{Introduction}
\label{sec:introduction}

The large-scale structure (LSS) of the Universe emerges from the gravitational growth of primordial density fluctuations, giving rise to a vast network of filaments, walls, clusters, and underdense regions known as cosmic voids \citep[e.g.,][]{Zeldovich1970, deLapparent1986, Bond1996, VAN_DE_WEYGAERT_2011, cautun_2014, kitaura_2020, Peebles1980, Aycoberry2024}.
Voids occupy most of the cosmic volume, while containing only a small fraction of the galaxy population, providing a unique laboratory for testing cosmological models and the physics of galaxy formation in low-density environments \citep{pan_2012, ricciardelli_2014}.
Their distinct dynamical conditions—such as low matter density, reduced gravitational interaction, and faster local expansion—strongly affect how galaxies evolve compared to those in denser environments \citep{ricciardelli_2014}.
Voids are therefore key to understanding the interplay between large-scale environment and galaxy properties.
Despite their significance, there is no unique or universal definition of a cosmic void.
Different algorithms identify underdense regions according to different criteria, resulting in catalogs with varying geometrical and dynamical characteristics \citep{colberg_2008, cautun_2018}.
Some methods define voids as spherical underdensities with a fixed threshold in the integrated density contrast \citep{padilla_2005, ruiz_void_2015}, while others rely on the topology of the density field using watershed techniques, such as the \zobov\  algorithm \citep[ZOnes Bordering On Voidness,][]{neyrinck_2008}.
Recently, alternative approaches have emerged, such as the \popcorn\ algorithm \citep{paz_2022}, also based on threshold in the integrated density contrast but with a complex approximation of the void's shape, and the \revolver\ framework \citep[REal-space VOid Locations from surVEy Reconstruction,][]{nadathur_2019}, which provides two watershed-driven void definitions that differ in the underlying tessellation: a cubic-voxel grid (\revolvervoxel) and a Voronoi tessellation (\revolverzobov).
These methodologies can yield substantially different catalogs even when applied to the same data set, highlighting the need for systematic comparisons across definitions.
Previous research has shown that the way halos and galaxies populate low-density regions is different from the cosmic average.
These variations are well understood regarding the absence of high-mass halos in voids \citep[e.g.,][]{Mo_White_1996, Sheth_Lemson_1999}.
However, results regarding an environmental dependence at fixed halo mass remain ambiguous.
While some studies find no significant dependence \citep[e.g.,][]{Abbas_Sheth_2007, paranjape_2018}, others report detectable variations \citep[e.g.,][]{Artale2018, Zehavi2018, bose_hod_2019}.
In particular, studies of halo occupation within spherical voids reveal that the halo occupation distribution (HOD) is significantly suppressed for halo masses above $10^{12} h^{-1}M_\odot$, depends on the density thresholds used to define the void boundaries, and becomes increasingly pronounced toward lower redshifts \citep{Alfaro2020, alfaro_hod_2022, Alfaro2025}.
This diversity of findings suggests that the inferred galaxy-halo connection may be sensitive to the specific definition of the environment, highlighting the critical importance of comparing results across different void-finding algorithms.
Against this backdrop, a complementary aspect in the study of cosmic environments is the concept of galaxy bias, which quantifies how the spatial distribution of luminous tracers deviates from that of the underlying dark matter field \citep[e.g.,][]{kaiser_1984, desjacques_2018}
Large-scale galaxy bias has been extensively characterized, with numerous works showing its dependence on global properties as well as on secondary halo characteristics linked to formation history \citep[e.g.,][]{Bardeen1986, Mo_White_1996, sheth_2004, Gao_2005, 
Wechsler2006, Gao2007, Tinker2010, Dalal2008, Lazeyras2017, Salcedo2018, Mao2018, SatoPolito2019, Contreras2019, Tucci2021, Contreras2021, MonteroDorta2021, Balaguera2024}.
More recently, several efforts have explored how galaxy bias relates to the geometry and dynamics of the cosmic web, identifying additional dependencies associated with large-scale structure \citep[e.g.,][]{MonteroDorta_Rodriguez_2024, Rodriguez_MonteroDorta2025}.
Within this broader context, the concept of individual galaxy bias ($b_i$) has been introduced to quantify the contribution of each galaxy to the total linear bias of its population \citep{paranjape_2018, han_2019, Balaguera2024, montero-dorta_2025}.
This framework facilitates a direct exploration of how the bias of a galaxy depends on its local and large-scale environment, providing a finer probe of the link between galaxy formation and the underlying matter distribution.
In particular, examining the behavior of $b_i$ within cosmic voids offers insight into how galaxies trace the matter field in the most extreme low-density regions, where the bias can even become negative.
Building on these ideas, \citet{montero-dorta_2025} formalized the notion of a void bias profile — the radial dependence of the mean galaxy bias when distances are normalized by the characteristic size of each void (e.g. $R/R_{\rm void}$). 
That work showed that the mean bias systematically varies with normalized void-centric distance and that the profile encodes environmental information (for example, differing behavior for voids embedded in overdense versus underdense surroundings). 
The term voids bias profile therefore provides a compact, physically motivated descriptor to discuss how galaxy–matter coupling changes across void interiors and boundaries, and it offers a natural target for testing whether this behavior is intrinsic to voids or contingent on the identification method.
Moreover, understanding the detailed bias behavior of galaxies and voids has broader cosmological significance, because void statistics and void–galaxy correlations have been shown to be sensitive probes of the underlying cosmological model and can be used to calibrate galaxy bias and improve cosmological parameter constraints from large-scale structure surveys \citep[e.g.][]{Hamaus2015, Contarini2019}.
In this work, we extend the results of \citet{montero-dorta_2025}, testing the robustness and possible universality of the bias profile of voids across several alternative definitions of voids.
Specifically, we analyze the voids identified with four distinct algorithms: \revolvervoxel, \revolverzobov, \zobov\ and \popcorn.
Our goal is to assess how the individual galaxy bias behaves within the void regions defined by each method and to evaluate the correspondence between galaxies with negative individual bias and the locations of voids across the different catalogs.
The use of multiple void definitions enables us to quantify the dependence of galaxy bias on the adopted void-finding methodology and to evaluate the robustness of previous results regarding the environmental modulation of bias. The analysis presented here is based on the same underlying data as in \citet{montero-dorta_2025}, based on the hydrodynamical cosmological simulation IllustrisTNG\footnote{\url{http://www.tng-project.org}}\citep{nelson_tng_2018, naiman_tng_2018, marinacci_tng_2018, springel_tng_2018, pillepich_tng_2018}.
This ensures full consistency with earlier studies while allowing for a direct extension of the methodology to alternative void identification schemes.
This paper is organized as follows.
In Section \ref{sec:data}, we describe the data sets of simulated galaxies used.
Section \ref{sec:voids_data} detailed all the void identification algorithms used and the final catalogs selected.
Section \ref{sec:bias_in_voids} presents the methodology for computing individual galaxy bias and the main results, in particular, the individual galaxy bias profile in each void definition.
Other important results are shown in Section \ref{sec:bias_with_delta}, where we explored the correlation of galaxies with negative values of individual bias with the density-threshold based voids definitions.
Finally, Section~\ref{sec:summary} summarizes our main conclusions and discusses the implications for future studies of galaxy bias in the cosmic web.
%

\section{Data}
\label{sec:data}
\subsection{The TNG300 simulation}
\label{sec:TNG300_sim}
We base our analysis on the $z=0$ snapshot of the IllustrisTNG simulation suite, specifically the largest box, TNG300-1 (hereafter TNG300).
This is a cosmological magneto-hydrodynamical simulation run with the moving-mesh code \textsc{arepo} \citep{springel_2010}, and it incorporates state-of-the-art subgrid models that account for star formation, stellar evolution, chemical enrichment, gas cooling, and feedback from both supernovae and active galactic nuclei (AGN).
TNG300 follows the evolution of dark matter and baryonic matter within a cubic comoving box of side length $205\hmpc$, making it ideal for studying large-scale environmental effects such as those associated with cosmic voids.
The simulation resolves the complex interplay between baryonic and dark matter components across a broad range of scales, from galactic to cosmological, and provides a robust statistical sample of galaxies suitable for the study of bias and environmental dependencies.
The adopted cosmological model corresponds to a flat $\Lambda$CDM cosmology consistent with the Planck 2016 results \citep{planck_2016}, defined by the following parameters: total matter density $\Omega_{\rm m} = 0.3089$, baryon density $\Omega_{\rm b} = 0.0486$, dark energy density $\Omega_{\Lambda} = 0.6911$, amplitude of matter fluctuations $\sigma_{8} = 0.8159$, Hubble parameter $h = 0.6774$, and scalar spectral index $n_{\rm s} = 0.9667$.
The mass resolution for dark matter particles is $m_{\mathrm{dm}} = 4.0 \times 10^{7}\hmsun$, while the initial baryonic mass resolution is $m_{\mathrm{gas}} = 7.6 \times 10^{6}\hmsun$.
The gravitational softening length for stellar particles is $\epsilon_{\mathrm{stars}} = 0.74\hkpc$, ensuring an adequate spatial resolution for both galaxy-scale and large-scale structure analyses.
%

\subsection{Galaxy Catalog Construction}
\label{sec:gal_catalog}

The galaxy sample used in this work is identical to that used in \citet{montero-dorta_2025}, allowing a direct comparison between our current results and previous analyses based on spherical voids.
Galaxies are identified as self-bound substructures with a stellar component using the \textsc{subfind} algorithm \citep{springel_2001, dolag_2009}, which operates on top of Friends-of-Friends \citep[FoF,][]{huchra_geller_fof} halos. 
We select all subhalos from the $z=0$ snapshot that satisfy the stellar mass threshold $\log_{10}(\mathrm{M_{*}}/\hmsun]) > 8.33$, corresponding to galaxies resolved with at least $\sim30$ stellar particles.
This conservative limit ensures that the resulting sample is not affected by incompleteness at the low-mass end while maintaining sufficient statistics for clustering measurements.
For each galaxy, we use the 3D comoving Cartesian coordinates and peculiar velocities, as well as the total stellar mass $M_{*}$.
Throughout the analysis, we restricted the study to central galaxies, those residing at the potential minimum of their host halos, to minimize the influence of satellite dynamics on the estimation of individual galaxy bias ($b_i$).
This selection provides a cleaner link between the galaxy position and the underlying dark matter distribution.
In our catalog, and following the prescription of \cite{Balaguera2024a}, the individual linear galaxy bias of a galaxy $i$ at position $\mathbf{r}$, $b_i$, was computed as  
\begin{equation}
    b_i =\frac{\sum_{j,k_{j}<k_{max}}N^{j}_{k}\langle \exp[-i\bf{k}\cdot \bf{r}_{i}]\delta_{\mathrm{DM}}^{*}(\bf{k}) \rangle_{k_{j}}}{\sum_{j,k_{j}<k_{max}} N^{j}_{k}P_{\rm DM}(k_{j})},
   \label{eq:bias}
\end{equation}
where $\delta_{\mathrm{DM}}^{*}(\mathbf{k})$ denotes the Fourier transform of the dark-matter density contrast, $P_{\rm DM}(k_j)$ is the matter power spectrum, and $N^{j}_{k}$ is the number of Fourier modes contained in the $j$$^{\rm th}$ spherical shell.
\footnote{Individual galaxy-bias assignments were carried out using the \texttt{CosmiCCcodes} library, \url{https://github.com/balaguera/CosmicCodes}} For an in-depth and intuitive explanation of the derivation of Eq.~\ref{eq:bias}, we refer the reader to \cite{paranjape_2018} and \cite{Paranjape_Alam_2020}. In essence, the method exploits the basic properties of discrete Fourier transforms to express the bias of a tracer population as the average of the individual contributions from each object.
The final catalog thus serves a dual purpose:
(i) it defines the tracer distribution used to construct the density field for cosmic void identification; and
(ii) it constitutes the sample for which we compute the large-scale individual galaxy bias, both globally and within the regions classified as voids by the different identification algorithms.
This consistent use of the same galaxy population for void detection and bias estimation ensures that the inferred environmental dependencies directly reflect intrinsic variations in the galaxy–matter connection rather than sample-selection effects.
%

\section{The void finders and their corresponding catalogs}
\label{sec:voids_data}

\subsection{Conceptual overview}
\label{sec:voids_overview}
The structural and dynamical properties of cosmic voids, such as their size distribution, topology, and enclosed mass, strongly depend on the identification algorithm adopted.
Since each method relies on different assumptions regarding geometry, density estimation, and boundary definition, cross-comparing their results is essential to assess the robustness of void-related statistics and to understand the environmental dependence of galaxy properties \citep{colberg_2008, cautun_2018}.
The baseline analysis presented in \citet{montero-dorta_2025} employed a spherical void finder implemented through the \sparkling\footnote{\url{https://gitlab.com/andresruiz/Sparkling}} algorithm \citep{ruiz_void_2015,ruiz_into_2019}, which identifies local minima in the galaxy density field and grows the largest non-overlapping spheres around them while maintaining an integrated density contrast below $\Delta_{\mathrm{lim}} = -0.9$.
Although conceptually simple and computationally efficient, this method restricts voids to spherical geometry, which may not fully capture the irregular shapes and hierarchical nature of true underdensities in the cosmic web \citep{ricciardelli_2014}.
To extend the analysis beyond the spherical approximation, we employ three advanced algorithms that represent complementary approaches to void identification: \zobov\footnote{\url{http://skysrv.pha.jhu.edu/~neyrinck/voboz}} \citep{neyrinck_2008}, \revolver\footnote{\url{https://github.com/seshnadathur/Revolver}} \citep{nadathur_2019} and \popcorn\footnote{\url{https://gitlab.com/dante.paz/popcorn_void_finder}} \citep{paz_2022}. 
\zobov\ provides a completely parameter-free and geometry-independent method based on the Voronoi Tessellation Field Estimator \citep[VTFE,][]{Schaap2007} and the watershed transforms \citep{Platen2007}, producing highly irregular structures that closely reflect the topology of the cosmic density field \citep{cautun_2018}.
\revolver\ builds on the framework \zobov\ by offering density estimation modes based on VTFE (\revolverzobov) and grid-based (\revolvervoxel), with the key advantage of allowing the derivation of spherical-equivalent voids from watershed structures for consistent comparisons using multiple methods. 
Finally, \popcorn\ represents a hybrid approach that overcomes void fragmentation by merging multiple overlapping spheres into extended, free-form structures that satisfy a global integrated underdensity constraint, preserving the physical interpretability of threshold-based methods while providing a more realistic description of void topology. 
The following subsections describe the technical implementation and resulting catalog characteristics for each algorithm, forming the basis for the comparative analysis of individual galaxy bias presented in the subsequent sections.
%

\subsection{\zobov\ voids}
\label{sec:zobov_data}

The \zobov\ analysis was performed using the \texttt{Void Finder Toolkit}\footnote{\url{https://github.com/FeD7791/voidFinderProject}} (VFT), applied to the $z=0$ TNG300 galaxy catalog described in Section~\ref{sec:gal_catalog}. 
The VFT is a free framework designed for comparing cosmic void catalogs.
It provides convenient wrappers for several void-finding algorithms (most notably \zobov\ ) and offers streamlined access to the tracers associated with each void. 
We adopt this feature for simplicity, as it allows us to avoid manually recovering tracers from the original \zobov\ output.
The \zobov\ algorithm identifies voids through a parameter-free topological approach that operates directly on a discrete distribution of tracers.
Its core procedure can be summarized in three main stages:
\begin{itemize}
    \item
    Local number densities are computed using the VTFE.  
    Each galaxy is assigned a Voronoi cell, and its local density $\rho$ is defined as the inverse of that cell's volume, $\rho \propto 1/V_{\text{cell}}$.
    \item
    The tessellated space is partitioned into contiguous regions, or \emph{zones}, each surrounding a local density minimum.  
    Galaxies iteratively ``flow'' along neighboring cells toward lower-density neighbors until they reach a minimum, defining the core of each zone.
    \item
    The zones are progressively merged according to the watershed analogy.  
    As the density threshold rises, neighboring zones are joined until the ``water'' spills into a zone with a lower minimum density.
    Thus, each final void is defined as the union of all zones connected before this flooding event occurs.
\end{itemize}
The initial application of \zobov\ yielded a raw catalog containing $1051$ structures.  
Since our objective is to connect individual negative galaxy bias values ($b_i < 0$) to well-defined low-density environments, a rigorous filtering procedure was applied to obtain a robust set of independent voids.
The filtering ensures that the final sample represents statistically significant, non-overlapping underdense regions suitable for large-scale structure analysis.
The following criteria were applied sequentially to construct the final catalog:
\begin{enumerate}
    \item 
    To keep some statistical significance on the catalog, we clean voids based on the Poisson probability of being Fake. To retain voids with the lowest chance of their density contrast arising from Poisson noise, voids with a $\sigma >= 3$ level of significance are optimal to be retained \citep{neyrinck_2008}. However, selecting such a value leaves us with approximately $2.4\%$ of the catalog (the complete sample was close to $1000$ voids), with 2-$\sigma$ we retain approximately $9.8\%$ of the catalog, and with 1-$\sigma$ 34\% of the catalog. In this case, too little of the catalog remains for performing the bias analysis (even for 2-$\sigma$). For 1-$\sigma$, we will have a bigger catalog but with roughly 30\% of the objects resulting from Poisson noise. This is a tradeoff we assume to keep the analysis somewhat plausible, instead of considering the whole catalog, which could lead to more misleading results.
    \item 
    As mentioned earlier, the \zobov\ algorithm merges \emph{zones} to form voids. However, this procedure can sometimes produce large voids that do not correspond to genuine underdensities, as shown in \cite{cai2017lensing}. To avoid this issue, we opt to treat each zone as an individual, non-overlapping void by definition.
    \item 
    To guarantee completeness and mitigate the effects of shot-noise dominance in low density and low scale regions, we kept only voids with effective radii $R_{\text{eff}} > 9\hmpc$, where $R_{\text{eff}}$ is the radius of a sphere with equivalent volume. This scale marks the transition in the void size function (VSF) where the slope becomes negative, indicating a robust detection regime above the resolution limit.
\end{enumerate}
After applying these criteria, the final catalog consists of $283$ voids.  
Their main statistical properties, including median radius and density contrast, are summarized in Table \ref{tab:voids_catalogs}.

\subsection{\revolver\ voids}
\label{sec:revolver_data}

The \revolver\ analysis was performed using the publicly available code, applied to the $z=0$ TNG300 galaxy catalog described in Section~\ref{sec:gal_catalog}.
\revolver\ is a sophisticated void-finding framework derived from watershed-based void identification algorithms.
Unlike purely spherical methods or standard watershed approaches, \revolver\ provides a hybrid methodology that bridges these two paradigms, offering both flexibility in void geometry and computational efficiency.
The \revolver\ algorithm operates in two distinct modes, each employing a different strategy to estimate the local tracer density field before applying the watershed transform:
\begin{itemize}
    \item 
    \revolverzobov: This mode follows the methodology of the original \zobov\ algorithm, VTFE to compute local densities from the discrete galaxy distribution.
    This tessellation-based approach provides an adaptive density estimate that is particularly robust to shot-noise effects, as the effective smoothing scale naturally varies with the local tracer density.
    \item 
    \revolvervoxel: In this mode, the density field is reconstructed using a particle-mesh interpolation scheme on a regular grid.
    Galaxy positions are interpolated onto the grid cells using a cloud-in-cell algorithm, and the resulting density field is normalized to the mean density of the simulation box.
\end{itemize}
Following density estimation, both modes apply a modified watershed algorithm to identify underdense regions.
Local density minima are identified as potential void centers, and the watershed basins surrounding these minima define the spatial extent of each void.
As in \zobov\, the boundaries between voids are determined by the ridges in the density field where the neighboring basins meet.
This topological approach naturally produces voids with irregular, non-spherical morphologies.
A crucial feature of \revolver\ for our analysis is its ability to derive spherically symmetric representations of these irregular watershed voids.
For each identified void, the algorithm calculates an effective spherical radius, $R_{\mathrm{eff}}$, where the total volume to be defined takes into account all Voronoi cells (in \revolverzobov) or grid cells (in \revolvervoxel) belonging to the void basin.
Additionally, \revolver\ determines an effective void center as the barycenter of the galaxies of the zone that defines each sphere.
The use of these spherically symmetric representations offers several advantages for our study of individual galaxy bias.
First, the effective radius $R_{\mathrm{eff}}$ provides a well-defined characteristic scale for each void that facilitates a direct comparison with the spherical void catalog while preserving information about the underdensity captured by the watershed algorithm.
Second, the effective center provides a natural reference point for measuring radial profiles and distances, which is essential for computing the bias profile as a function of position within and around voids.
Third, this approach represents an intermediate strategy between purely spherical methods (which impose a strict density constraint) and purely watershed methods (which may produce highly fragmented or hierarchical structures that are difficult to statistically analyze).
The initial application of \revolver\ to the TNG300 galaxy catalog yielded $604$ voids using the \revolverzobov\ mode and $2822$ voids using the \revolvervoxel\ mode.
However, not all identified structures are suitable for our analysis of individual galaxy bias.
It is important to note that the spherical representations derived by \revolver\ can exhibit significant spatial overlap.
Unlike the strictly non-overlapping spheres produced by traditional spherical void finders, the effective spheres in \revolver\ correspond to the volumes of topologically distinct watershed basins, which may be adjacent or even nested within larger structures.
To construct robust final catalogs, we applied a series of filtering criteria designed to: (i) ensure completeness above the resolution limit, (ii) remove redundant detections caused by void overlap, and (iii) select physically meaningful underdensities.
The filtering procedure consisted of the following steps:
\begin{enumerate}
    \item 
    We applied a minimum radius threshold to exclude voids below the reliable detection limit.
    For \revolverzobov\ voids, we imposed $R_{\mathrm{eff}} > 10\hmpc$, while for \revolvervoxel\ voids, we used $R_{\mathrm{eff}} > 7.5\hmpc$.
    Again, these thresholds were chosen considering in both cases the scale where the VSF slope becomes negative and mark the transition of the function into a robust power-law regime where spurious detections due to Poisson noise are negligible.
    The slightly lower threshold for voxel voids reflects the different noise properties of grid-based density estimation compared to tessellation-based methods.
    \item 
    The overlapping nature of \revolver\ effective spheres necessitates a systematic procedure for identifying and resolving redundant detections.
    We first classified all voids by the effective radius $R_{\mathrm{eff}}$ in descending order.
    Starting with the largest void, we identified all voids whose effective spheres overlap with it by computing pairwise center-to-center distances and comparing them to the sum of the effective radii.
    For each pair overlapping $(i,j)$ with $R_{\mathrm{eff}, i} > R_{\mathrm{eff},j}$, we ordered the pair by the average density contrast of the smaller void $\overline{\Delta}_j = \Delta_j/V_j$, where $\Delta_j$ and $V_j$ are the integrated density contrast and volume of the void.
    If the larger void satisfied $\overline{\Delta}_i > 0$ (overdense), the smaller void satisfied $\overline{\Delta}_j < 0$ (underdense), and the absolute difference $|\overline{\Delta}_i - \overline{\Delta}_j| > 0.5$, then we retained the smaller void and removed the larger one from the catalog.
    This criterion identifies cases where the effective sphere of a larger structure incorrectly encompasses a genuine smaller void with a significantly deeper density minimum.
    Otherwise, we retained the larger void and removed the smaller one, under the assumption that the smaller structure represents a substructure or boundary feature of the larger void rather than an independent underdensity.
    This process was repeated iteratively for all overlapping pairs in order of decreasing $\overline{\Delta}_j$ of the smaller void.
\end{enumerate}

After applying these selection criteria, the final \revolver\ void catalogs consist of $153$ voids identified in the \revolverzobov\ mode and $402$ voids identified in the \revolvervoxel\ mode.
The main statistical properties of these catalogs are summarized in Table \ref{tab:voids_catalogs}.
These catalogs provide a complementary perspective to the spherical and \zobov\ void samples, allowing us to assess how the identification methodology affects their relationship to the spatial distribution of galaxies with negative individual bias.
%

\subsection{\popcorn\ voids}
\label{sec:popcorn_data}

The identification of voids based on a fixed integrated density contrast was performed using the \popcorn\ void finder.
This method defines voids as free-form, non-spherical structures constructed as the union of maximal spheres that jointly satisfy a prescribed integrated underdensity threshold.
The algorithm generalizes the spherical void finder strategy by iteratively merging overlapping spheres, enabling the recovery of the full extent of underdense regions and yielding a more accurate, continuous representation of void boundaries.
Following the procedure described in \cite{paz_2022}, the algorithm proceeds through the following main steps:
\begin{itemize}
    \item 
    The surface of each initial spherical void is uniformly covered with seed points using a Fibonacci double-spiral distribution.
    \item 
    Each seed is expanded to the largest possible radius for which the integrated density contrast of the combined region satisfies the condition $\Delta < \Delta_{\rm lim}$. Where $\Delta_{\rm lim}$ is a predefined threshold to keep the density contrast below this value (commonly $-0.9$ or $-0.8$).
    \item 
    Spheres with radii smaller than a specified minimum threshold are discarded.
    \item 
    The largest valid seed is accepted and incorporated into the void structure, after which the surface is reseeded to allow further expansion.
    \item 
    The process is repeated iteratively until no additional spheres can be added, after which overlapping \popcorn\ voids are removed, retaining only those derived from the largest parent spherical void.
\end{itemize}
For consistency with the fiducial spherical void catalog of \citet{montero-dorta_2025}, we adopt an integrated density contrast threshold of $\Delta_{\rm lim} = -0.9$.
This threshold corresponds to regions where the integrated density is at most $10\%$ of the mean tracer density, a value commonly adopted in spherical void analyses.
The centers of \popcorn\ voids are defined like the center of the initial spherical void, and its effective radius, $R_{\mathrm{eff}}$, is given by the equivalent sphere with the same center and a total comoving volume of the union of all its constituent spheres.
The initial application of the \popcorn\ algorithm to our TNG300 galaxy sample produced a raw catalog containing $2529$ candidate voids.
To ensure the physical reliability of the identified structures and minimize the contamination by shot-noise, we applied a selection based on the radius of shot-noise ($R_{\mathrm{shot}}$), as defined in \citet{paz_2022}.
This scale represents the typical radius at which a Poisson sampling of the tracer population would, on average, enclose one galaxy.
It thus provides an empirical lower limit to the physical scale at which voids can be reliably resolved, given the sampling density.
In our sample, only the voids that satisfy $R_{\mathrm{eff}} > R_{\mathrm{shot}} = 6.5\hmpc$ are retained in the final catalog.
This cutoff efficiently removes spurious detections dominated by shot noise and avoids the artificial fragmentation of small underdense regions.
After applying this criterion, the final \popcorn\ catalog comprises $495$ statistically robust voids, representing physically meaningful underdensities in the TNG300 galaxy distribution.
The main statistical properties of this catalog, including median radius and density contrast, are summarized in Table \ref{tab:voids_catalogs}.
\\\\

\begin{figure}[ht!]
\begin{center}
\includegraphics[width=0.9\columnwidth]{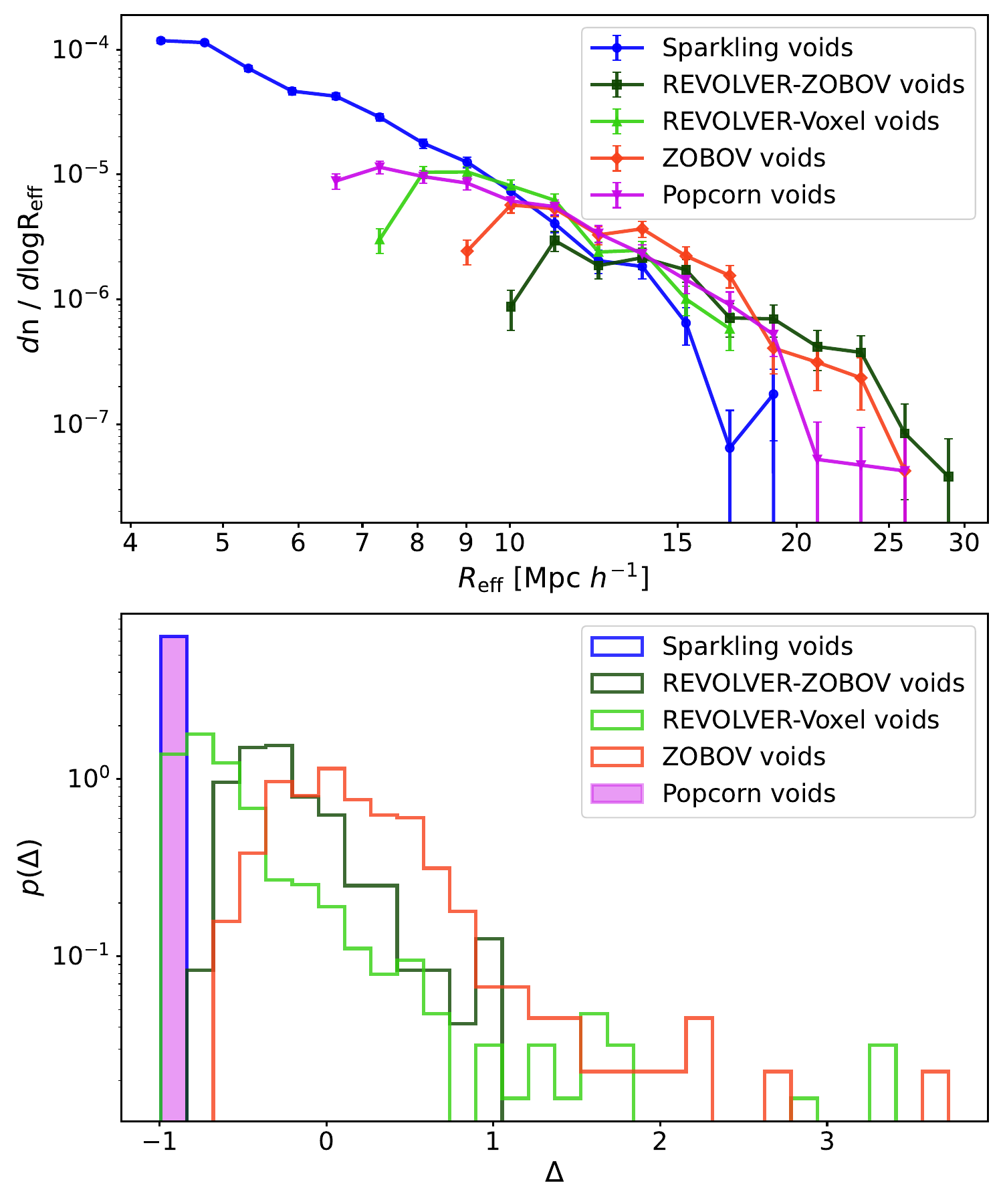}
\end{center}
\caption{\label{fig:voids_prop} 
Comparison of the five void catalogs used in this work. The top panel shows the void size function for each voids catalog. The bottom panel shows the normalized distribution of the integrated density contrast, $\Delta$, for these same catalogs.}
\end{figure}

The deployment of a suite of diverse void identification algorithms (\sparkling, \zobov, \revolvervoxel, \revolverzobov\ and \popcorn) necessarily yields catalogs exhibiting intrinsic structural variability. 
These catalogs demonstrate differences in size, integrated density profile, and resulting morphology. 
This comprehensive methodological comparison allows us to assess the stability and environmental dependence of galaxy properties, such as the individual galaxy bias, across distinct geometrical and structural void definitions.
The quantitative summary of the key properties of the derived void populations is provided in Table \ref{tab:voids_catalogs} and Figure \ref{fig:voids_prop}.
Table \ref{tab:voids_catalogs} lists the major statistical characteristics of the final void samples, including the total number of identified structures ($N_{\mathrm{void}}$), the cumulative volume fraction occupied by the voids ($Total_{\mathrm{vol}}$  in $\%$ of the box volume), the total number of galaxies residing within them ($N_{\mathrm{void \ galaxies}}$), and the full range and mean of the effective radii ($R_{\mathrm{eff, \ min}}$, $R_{\mathrm{eff, \ max}}$ and $R_{\mathrm{eff, \ mean}}$ in $h^{-1}\mathrm{Mpc}$) and integrated underdensities ($\Delta_{\mathrm{min}}$, $\Delta_{\mathrm{max}}$ and $\Delta_{\mathrm{mean}}$). 
%

\begin{table*}
    \caption{Main characteristics of the final void catalogs. The \sparkling\ catalog is the same one used in \citet{montero-dorta_2025}.}
    \label{tab:voids_catalogs}
    \centering
    {
    \scriptsize
    \begin{tabular}{c c c c c c c c c c}    
    \hline\hline 
    Void catalog & $N_{\mathrm{void}}$ & $N_{\mathrm{void \ galaxies}}$ & $Total_{\mathrm{vol}}$ & $R_{\mathrm{eff, \ min}}$ & $R_{\mathrm{eff, \ mean}}$ & $R_{\mathrm{eff, \ max}}$ & $\Delta_{\mathrm{min}}$ & $\Delta_{\mathrm{mean}}$ & $\Delta_{\mathrm{max}}$ \\
    \ & & & [$\%$]  & [$h^{-1}\mathrm{Mpc}$] & [$h^{-1}\mathrm{Mpc}$] & [$h^{-1}\mathrm{Mpc}$] & & & \\ 
    \hline\hline
    \sparkling\ & 2394 & 15030 & 37.20 & 4.07 & 6.09 & 19.68 & -0.95 & -0.91 & -0.90 \\
    \revolvervoxel\ & 401 & 54336 & 24.28 & 7.51 & 10.28 & 17.78 & -0.99 & -0.51 & 3.33 \\
    \revolverzobov\ & 152 & 84543 & 29.82 & 10.08 & 14.87 & 30.38 & -0.69 & -0.22 & 0.99 \\
    \zobov\ & 283 & 164009 & 36.37 & 9.03 & 13.00 & 25.25 & -0.59 & 0.17 & 3.73 \\
    \popcorn\ & 495 & 13411 & 31.51 & 6.51 & 9.97 & 25.11 & -0.92 & -0.90 & -0.90\\
    \hline
    \end{tabular}
    }
  \tablefoot{%
    Column 1: final void catalog; 
    Column 2: total number of identified structures; 
    Column 3: total number of galaxies residing within identified voids; 
    Column 4: cumulative volume fraction occupied by the voids; 
    Column 5: minimum value of $R_{\mathrm{eff}}$ by the voids sample; 
    Column 6: mean value of $R_{\mathrm{eff}}$ by the voids sample; 
    Column 7: maximum value of $R_{\mathrm{eff}}$ by the voids sample;
    Column 8: minimum value of $\Delta$ by the voids sample; 
    Column 9: mean value of $\Delta$ by the voids sample; 
    Column 10: maximum value of $\Delta$ by the voids sample.    
  }    
\end{table*}

%
Figure \ref{fig:voids_prop} summarizes the distribution functions of the large-scale properties for all five resulting void definitions. 
The upper panel shows the VSF, which shows the comoving number density of voids per logarithmic radius bin as a function of their equivalent radius ($R_{\mathrm{eff}}$, calculated from the volume of the void object). 
We utilize different color coding for each methodology: blue for the original \sparkling\ spherical voids of \cite{montero-dorta_2025}, dark green for \revolverzobov, light green for \revolvervoxel, orange for standard \zobov\ and violet for \popcorn\ voids.
This pattern of color will be maintained throughout all this work.
These VSFs exhibit significant divergences, reflecting the inherent differences between methods.
None of our catalogs reaches the smallest void radii included in the fiducial spherical sample of \citet{montero-dorta_2025}.
The \revolvervoxel\ catalog exhibits the narrowest radius distribution, indicating a stronger suppression of small and large voids, whereas both \revolverzobov\ and \zobov\ tend to identify the largest voids in the sample.
As expected, the \popcorn\ catalog mitigates artificial fragmentation of spherical voids at large scales \citep{paz_2022}, although it results in fewer very large objects than those detected by the watershed--based methods.
The bottom panel of Figure \ref{fig:voids_prop} illustrates the distribution of the integrated density contrast for these same catalogs. 
Although the \sparkling\ and \popcorn\ distributions are completely constrained in one single bin around $\Delta_{\mathrm{lim}} =-0.9$ by design, the watershed methods (\zobov\ and \revolver) naturally display greater variation in its internal density, underscoring their ability to encapsulate a wider range of core and boundary densities within their defined structures.
In general, \revolvervoxel\ voids exhibit lower density contrasts than \revolverzobov\ voids, while our \zobov\ void sample shows the highest density values among the five samples. 
%

\begin{figure*}[h!]
\begin{center}
\includegraphics[width=0.95\textwidth]{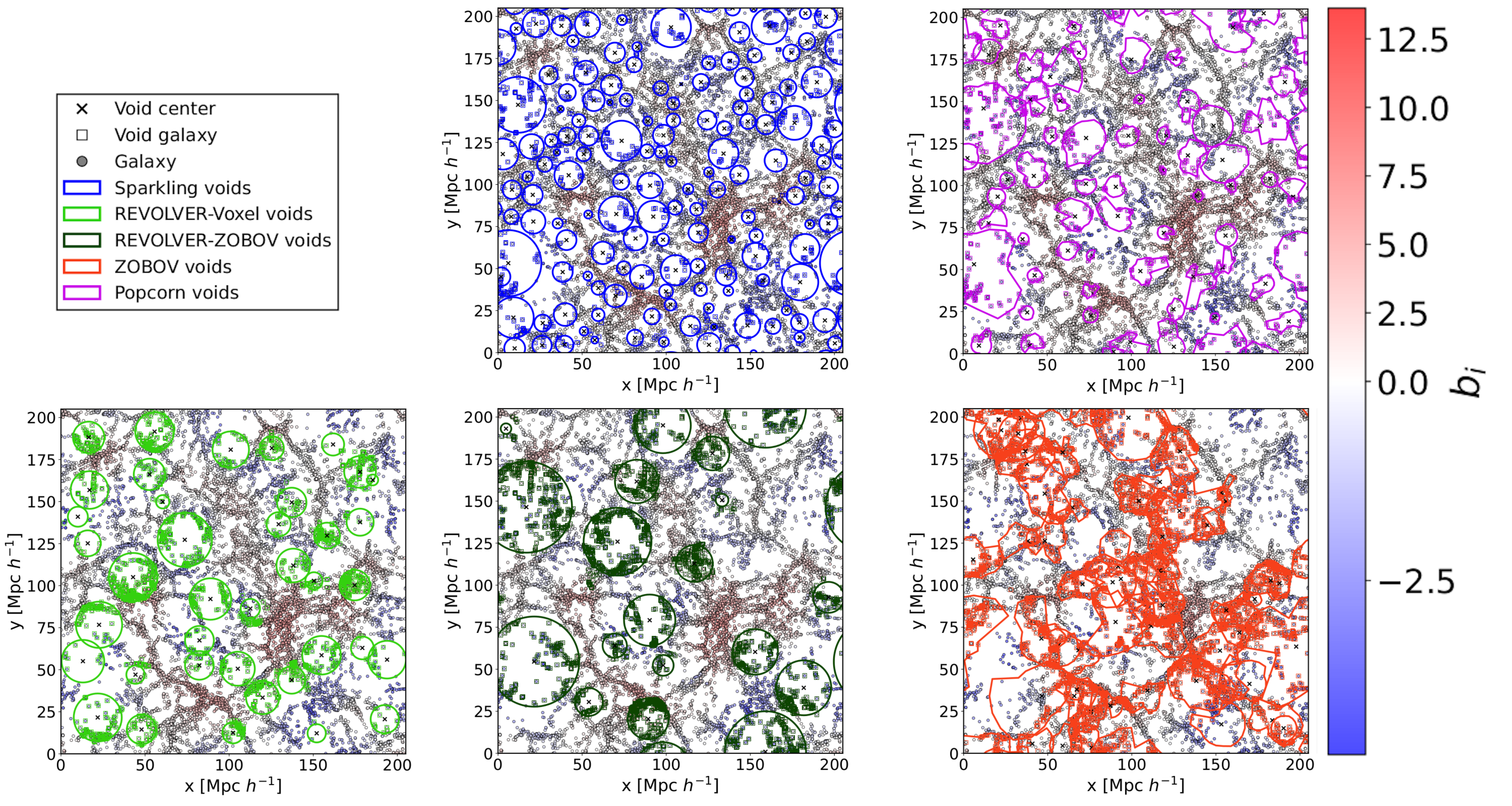}
\end{center}
\caption{\label{fig:voids_scatter}
Example slice of the TNG300 simulation box ($10\hmpc < z < 20\hmpc$) illustrating the galaxies belonging to voids (squares) and the boundaries that define them are colored by the five distinct void finders. Galaxy positions (scatter points) are color-coded by their individual large-scale galaxy bias, ranging from blue (lowest bias) to red (highest bias), like shows in the color bar.}
\end{figure*}

%
Figure \ref{fig:voids_scatter} provides a qualitative, yet crucial, visual comparison of the results, presenting a slice ($10\hmpc < z < 20\hmpc$) of the TNG300 catalog (see Section \ref{sec:gal_catalog}), demonstrating the diverse regions carved out by each finder (using the same color coding defined for Figure \ref{fig:voids_prop}).
To approximate the boundary of irregular voids (\zobov\ and \popcorn), we adopt a radial–angular sampling approach: from the void center, we divide the full solid angle into $20^{\circ}$ angular sectors and, within each sector, identify the galaxy farthest from the void center.
Connecting these outermost points yields a polygonal approximation of the contour of the void.
The apparent overlap between regions in the image arises from projection effects, since the figure shows a 2D slice of the 3D simulation volume.
This visualization confirms the high degree of morphological diversity and boundary complexity introduced by each technique. 
Critically, the scatter points representing the positions of individual galaxies within this slice are color-coded according to the value of their individual large-scale galaxy bias ($b_i$), ranging from low (blue) to high (red) bias. 
This map visually connects the structural properties of each void definition to the clustering properties of the galaxies contained within them, supporting the finding that galaxy bias is sensitive to the location and environment, even inside underdense regions.
%

\section{The galaxy bias profile in different void catalogs}
\label{sec:bias_in_voids}

This section presents our main results, namely, the dependence of the individual large-scale galaxy bias $b_i$ on the choice of cosmic void definition. 
We examine galaxies within underdense regions defined via five void finding algorithms defined in the previous section: the \sparkling\ spherical voids adopted by \citet{montero-dorta_2025}, the watershed-based \zobov\ method, the two variants of the \revolver\ catalogs (\textsc{voxel} and \zobov\ modes), and the free‐form integrated‐density approach \popcorn.
By comparing these distinct void definitions, we quantify how robust the measured bias profiles are with respect to the underdense‐region delineation.
We compute individual galaxy bias on an object‐by‐object basis, following the approach described by \citet{montero-dorta_2025}, i.e., the effective large‐scale bias of each galaxy is derived from its correlation with the dark‐matter density field on linear scales ($k \le 0.2 \, h \, \text{Mpc}^{-1}$).
This enables us to study the contribution of each galaxy $i$ to the overall clustering of the population, and in particular to investigate how $b_i$ depends on the environment of galaxies inside voids.
When averaged over a subsample, $\langle b_i \rangle$ provides an estimate of the mean tracer bias of that subsample, as inferred from the individual–based method. 
Under this framework, the average over a galaxy subset provides the effective large-scale bias as obtained from a two-point statistics analysis. 
This quantity approximates how well galaxies trace the underlying matter on large scales, although it does not strictly correspond to the traditional large-scale bias \citep[e.g.][]{Balaguera2024, montero-dorta_2025}.
%

\subsection{Distribution of Individual Galaxy Bias in Void Catalogs}
\label{sec:bias_distribution}

\begin{figure}[ht!]
\begin{center}
\includegraphics[width=0.9\columnwidth]{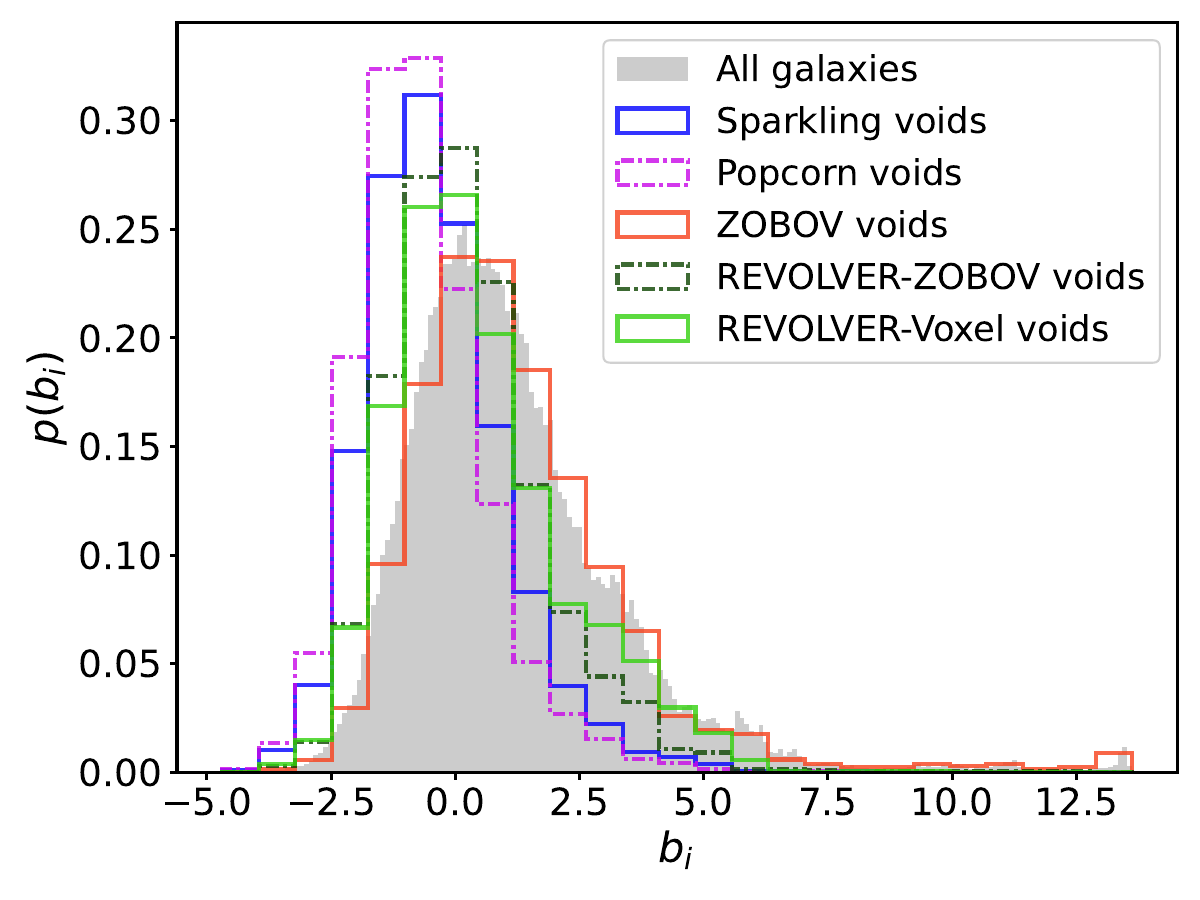}
\end{center}
\caption{\label{fig:bias_hist} 
Comparison of the individual bias distribution in galaxies of the five void catalogs used in this work, as indicated in the key figure. The gray shaded region represent the $b_i$ distribution on the complete TNG300 galaxies catalog.}
\end{figure}

Figure \ref{fig:bias_hist} shows the normalized distribution of $b_i$ for the full galaxy sample (gray shaded region) alongside the distributions for galaxies residing within the voids of each algorithm, as indicated in the key figure.
The full sample distribution gives our baseline mean large‐scale bias of the population (see Section \ref{sec:gal_catalog}).
We observe that galaxies located within underdense regions (for the \sparkling, \revolver, and \popcorn\ voids) generally show distributions shifted toward negative $b_i$ values.
In contrast, galaxies within \zobov-void regions show a $b_i$ distribution much closer to that of the overall sample. 
This finding implies that galaxies within strictly underdense voids tend to be anti-biased tracers of the matter field (they cluster less strongly than the dark matter), a result consistent with previous work on void bias profiles \citep[e.g.,][]{montero-dorta_2025}.
The differences in shape among the catalog‐specific $b_i$ distributions reflect the varying performance of each algorithm in isolating underdense zones versus including boundary or filamentary material where the galaxy bias is higher.
For example, the relatively “weak” shift of the \zobov\ distribution toward negative values suggests that the definition of the \zobov\ voids may include galaxies in regions less deeply underdense, thus diluting the anti-bias signal.
%

\subsection{Bias Profiles: Physical and Normalized Distances}
\label{sec:bias_profile}

\begin{figure}[ht!]
\begin{center}
\includegraphics[width=0.9\columnwidth]{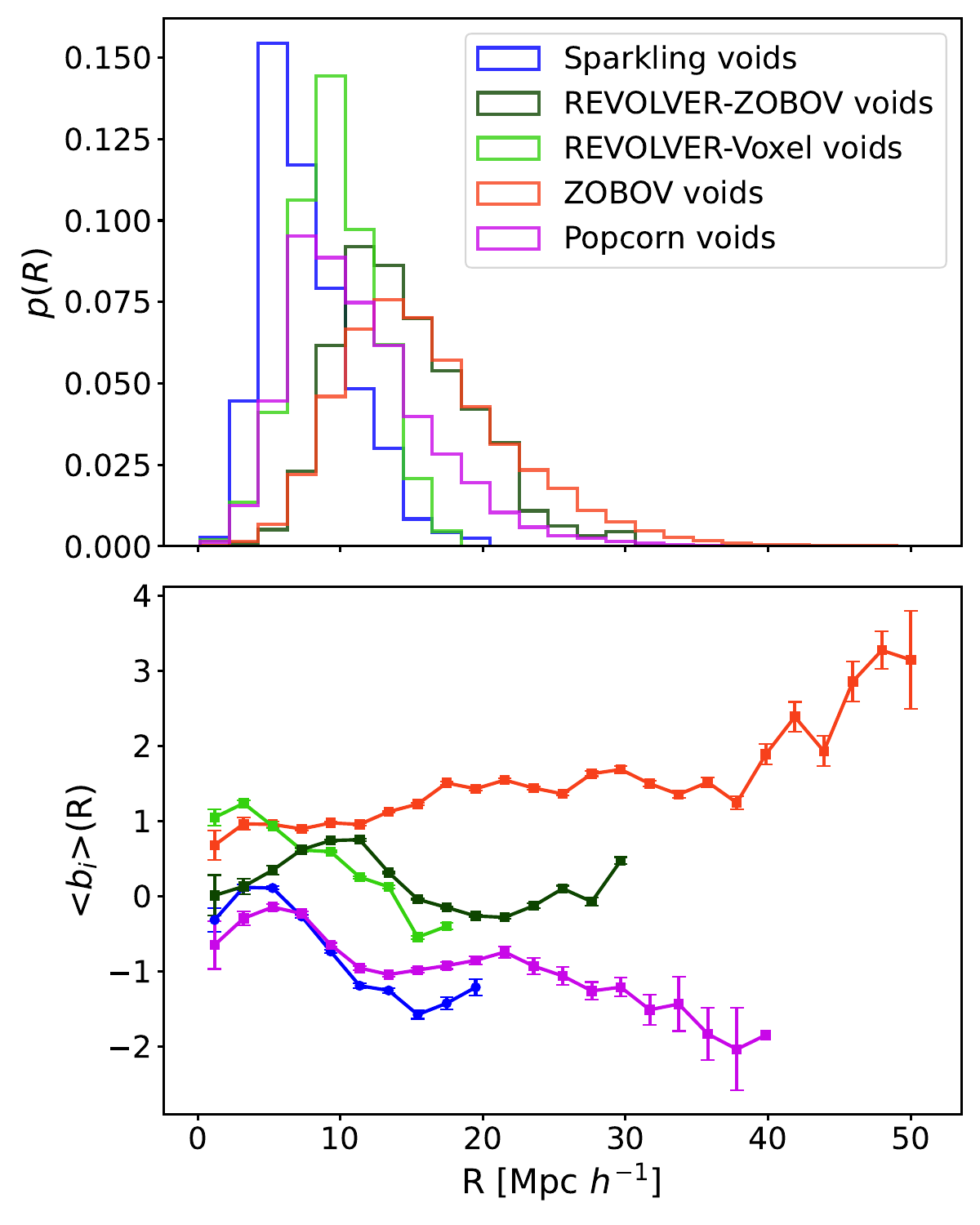}
\end{center}
\caption{\label{fig:linear_bias_profile} 
Relationship between the distance $R$ of voids galaxies to their corresponding center and $\langle b_i \rangle$. \textit{Top panel}: normalized distribution of $R$ for each void catalog. \textit{Bottom panel}: $\langle b_i \rangle$ profile in function of $R$ for each void catalog.}
\end{figure}

We now examine how the average individual bias $\langle b_i \rangle$ varies as a function of the distance $R$ of a galaxy from the void center and as a function of a normalized distance $R/R_{\rm max}$.
Here $R$ is measured in comoving $\hmpc$, and $R_{\rm max}$ denotes the distance from the center of the void to the most distant galaxy belonging to the void (this distance is not equivalent to the $R_{\mathrm{eff}}$ of each void, mainly in not spherical definitions).
The upper panel of Figure \ref{fig:linear_bias_profile} presents the distribution of physical distances $R$ for galaxies in each void catalog.
The VSF differs significantly between the algorithms (see Figure \ref{fig:voids_prop}), which complicates the direct comparison of $\langle b_i \rangle$ as a function of $R$. 
Nevertheless, the lower panel of Figure \ref{fig:linear_bias_profile} shows $\langle b_i \rangle$ versus $R$ (in $h^{-1}\mathrm{Mpc}$) to illustrate the trends in physical units.
We find that the \sparkling\ (blue) and \popcorn\ (violet) catalogs, which enforce a strict integrated underdensity threshold, exhibit nearly uniformly negative $\langle b_i \rangle$ across all distances, unlike the watershed catalogs.
In particular, for \popcorn, \sparkling\ and \revolvervoxel\ (light green) the bias trend to decreases as $R$ increases (i.e., to the largest voids).
In contrast, the \zobov\ (orange) and \revolverzobov\ (dark green) catalogs show a plane trend of $\langle b_i \rangle$.
Even the profile in \zobov\ voids increased with the highest $R$ values, indicating that galaxies near the edges in that catalog are less underdense and more biased.
%

\begin{figure}[ht!]
\begin{center}
\includegraphics[width=0.9\columnwidth]{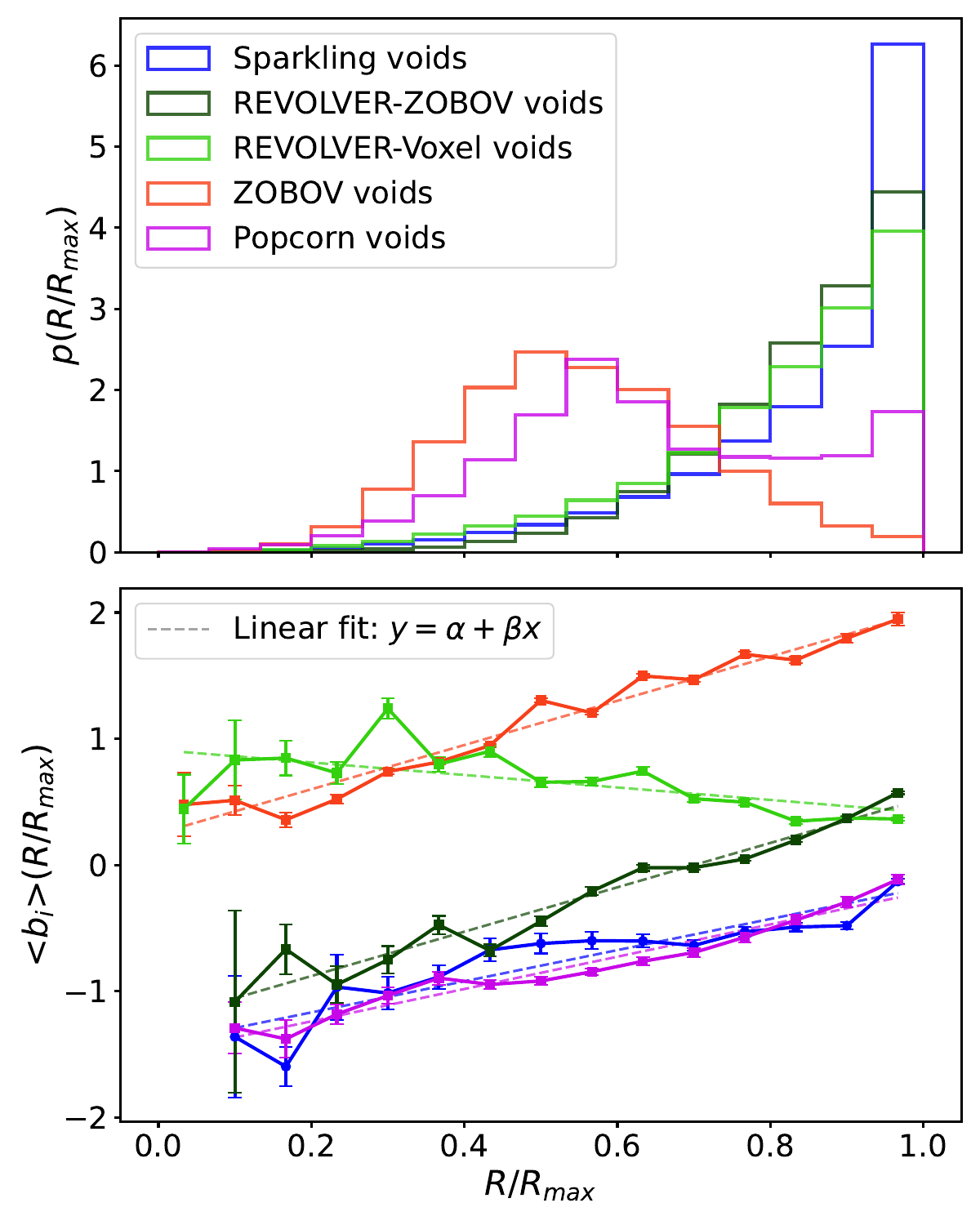}
\end{center}
\caption{\label{fig:normalized_bias_profile} 
Relationship between the normalized distance $R/R_{\rm max}$ of voids galaxies to their corresponding center and $\langle b_i \rangle$, where $R_{\rm max}$ represents the highest distance of a tracer to the center of their host void. \textit{Top panel}: normalized distribution of $R/R_{\rm max}$ for each void catalog. \textit{Bottom panel}: $\langle b_i \rangle$ profile in function of $R/R_{\rm max}$ for each void catalog. The dotted lines show the linear fit corresponding to each sample.}
\end{figure}

%
Because voids vary greatly in size and shape, we normalize the distance by dividing $R$ by $R_{\rm max}$ and plot the normalized distribution in the upper panel of Figure \ref{fig:normalized_bias_profile}.
This normalization enables a fair comparison of bias-profiles across voids of different sizes and extents.
For geometrically spherical voids (\sparkling\ in blue and \revolver\ family in green), the galaxy counts rise toward the void wall.
We note that this increase in galaxy counts towards the void boundaries (higher $R/R_{max}$) observed for spherical-like definitions is consistent with a geometric effect, as the volume of the spherical shells naturally increases with the distance from the center.
For \zobov\ (orange), the galaxy distribution peaks at intermediate $R/R_{\rm max}$, while in \popcorn\ (violet) we observe a bimodal distribution (one peak near the center, another near the boundary) due to its geometry resulting from the union of spheres.
In the normalized distance bias profiles (shown in the bottom panel of Figure \ref{fig:normalized_bias_profile}), with the exception of the \revolvervoxel\ catalog, all methods show the steepest (most negative) $\langle b_i \rangle$ near the center of the voids and a higher bias (less negative or even positive) near the walls.
We performed a linear adjustment for each profile represented by the dotted lines; the corresponding values of their parameters are shown in Table \ref{tab:linear_fit}. 
The \sparkling\ and \popcorn\ catalogs show very similar profiles and consistently the lower average bias across all distances and the lowest slopes, reflecting their strict underdensity threshold of $\Delta_{\rm lim} = -0.9$ imposed by construction. 
The \revolverzobov\ catalog has a steeper slope and even crosses into positive $b_i$ near the void rim.
The \zobov\ profile remains positive throughout the interior of the voids, but its slope is practically equal to the \revolverzobov\ slope.
These patterns suggest that watershed‐-based definitions (\zobov\ and \revolver) incorporate galaxies closer to denser filaments/walls (hence higher bias) than the strict density thresholds imposed in \sparkling\ and \popcorn.
Finally, the \revolvervoxel\ profile stands out for its negative slope.
This behavior can be attributed to the generally smaller sizes of these voids compared to the other watershed-based samples (see Figure \ref{fig:voids_prop}).
As a consequence, galaxies located near the walls of \revolvervoxel\ voids are likely to lie deeper within the interiors of the larger \revolverzobov\ or \zobov\ voids, where the $b_i$ values begin to increase.
This geometric effect naturally results in a systematically lower normalized profile for the \revolvervoxel\ sample.
Nevertheless, despite differences in amplitude and slope, the general shape of the bias profile, that increasing bias with increasing distance toward the void wall, holds across most of the methods.
This is in line with previous findings for voids identified in spherical definitions \citep{montero-dorta_2025}.
%

\begin{table}
    \caption{Linear fit parameters of the normalized void galaxies bias profile}
    \label{tab:linear_fit}
    \centering
    \scriptsize 
    \begin{tabular*}{0.9\columnwidth}{@{\extracolsep{\fill}}lcccc@{}}    
    \hline\hline 
    Void catalog & $\alpha$ & $\alpha$ Std Err & $\beta$ & $\beta$ Std Err \\ 
    \hline\hline
    \sparkling\ & -1.415 & 0.099 & 1.234 & 0.167 \\
    \revolvervoxel\ & 0.911 & 0.106 & -0.495 & 0.184 \\
    \revolverzobov\ & -1.231 & 0.074 & 1.757 & 0.123 \\
    \zobov\ & 0.249 & 0.058 & 1.752 & 0.101 \\
    \popcorn\ & -1.495 & 0.051 & 1.279 & 0.086 \\
    \hline
    \end{tabular*}
    \tablefoot{%
        Column 1: void catalog; 
        Column 2: $\alpha$ value (y-intercept); 
        Column 3: $\alpha$ standard error; 
        Column 4: $\beta$ value (slope); 
        Column 5: $\beta$ standard error; 
    }    
\end{table}

\subsection{Correlation of Individual Bias with Void Residency}
\label{sec:bias_matrix}

\begin{figure*}[h!]
\begin{center}
\includegraphics[width=0.9\textwidth]{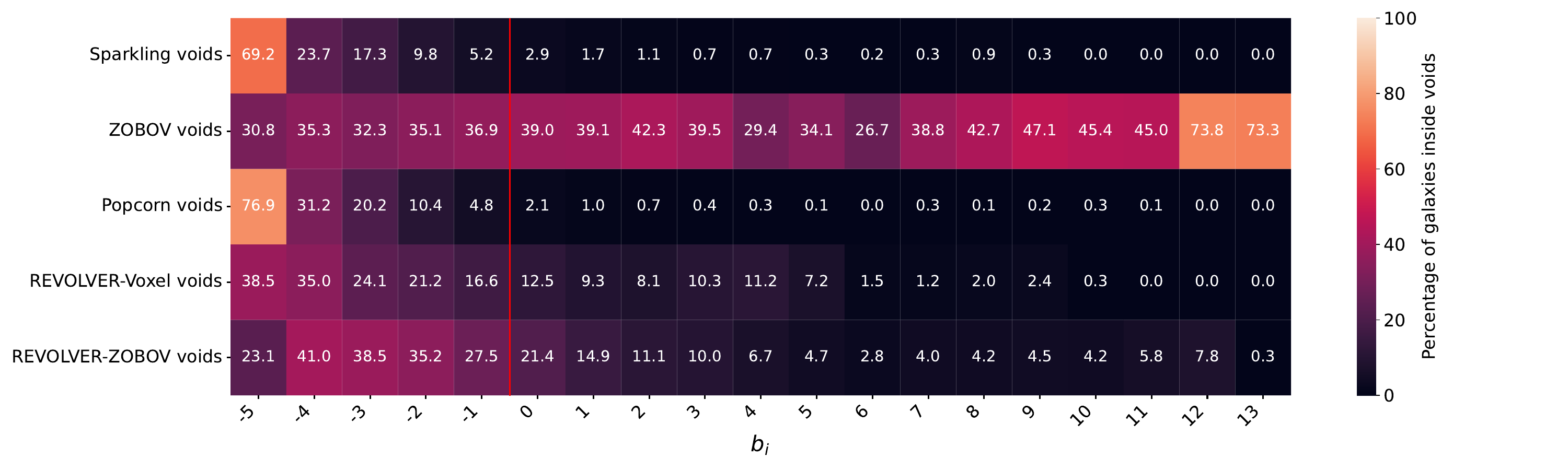}
\end{center}
\caption{\label{fig:bias_matrix}
Correlation matrix between the different void catalogs and $b_i$. Each cell gives the fraction of galaxies in the corresponding $b_i$-bin that lie within each void type. The colors are given by the lateral bar, and the red line divides the negative $b_i$-bins from other values.}
\end{figure*}

To further quantify the link between individual galaxy bias and void membership, we bin galaxies in the full TNG300 catalog (see Section \ref{sec:gal_catalog}) by $b_i$ and compute, for each bias bin and each void catalog, the fraction of galaxies that reside in voids of that catalog. 
Figure \ref{fig:bias_matrix} shows this as a correlation matrix where the $y$-axis lists the different catalogs of voids, the $x$-axis shows the bins in $b_i$, and each cell gives the fraction of galaxies in that $b_i$-bin that lie within voids of that type.
For the \sparkling, \revolvervoxel, \revolverzobov\ and \popcorn\ catalogs, we observe a clear correlation: the lower (more negative) the galaxy’s $b_i$, the higher the probability that the galaxy resides in the void catalog.
The \sparkling\ and \popcorn\ catalogs (density-threshold methods) especially show large occupancy fractions for $b_i < 0$ galaxies and very low fractions for $b_i > 0$.
By contrast, both \revolver\ catalogs also include a non-negligible fraction of high-bias galaxies ($b_i >2.5$), indicating that these voids include galaxies closer to denser structures. 
The \zobov\ catalog stands out as the most different: there is no strong correlation at low $b_i$, but rather a higher fraction of void galaxies in the highest $b_i$ bins.
This again supports the interpretation that \zobov\ voids include more boundary or wall galaxies, hence lifting the bias distribution toward positive values.
At the same time, when restricting the analysis to galaxies in fixed $b_i$ bins, particularly the most negative values, we find that these anti-biased galaxies populate \sparkling\ and \popcorn\ voids more frequently than \revolverzobov\ voids. 
This demonstrates that watershed-based voids are not simply the same deep underdense cores selected by density-threshold methods plus their surrounding walls; rather, they tend to trace shallower underdensities on average, while density-threshold finders preferentially pick out the deepest troughs where the most negatively biased galaxies reside.
%

\section{The dependence of galaxy bias on the void density threshold}
\label{sec:bias_with_delta}

In the previous section, we showed that void definitions based on an integrated density threshold (\sparkling\ and \popcorn) preferentially select galaxies with negative individual bias ($b_i<0$).
The fiducial threshold $\Delta_{\rm lim}=-0.9$ is a relatively restrictive value, so it is important to verify whether the observed correlation strengthens or weakens when the threshold is relaxed.
In this section, we therefore explore how the association between $b_i$ and void membership changes when the integrated underdensity criterion is progressively less stringent.
To perform this test, we re-identified voids in the same TNG300 tracer sample (Section~\ref{sec:gal_catalog}) using two additional integrated underdensity thresholds, $\Delta_{\rm lim} = -0.8$ and $-0.7$, for both the \sparkling\ and \popcorn\ finders. 
To preserve comparability with the fiducial catalogs ($\Delta_{\rm lim}=-0.9$), we applied the same completeness and shot-noise cuts used in the main analysis.
So, for the \sparkling\ catalogs, we retained only the voids with $R_{\rm void}>5\hmpc$ in $\Delta_{\rm lim}=-0.8$ and $R_{\rm void}>6\hmpc$ in $\Delta_{\rm lim}=-0.7$. 
For \popcorn\ we estimated the shot-noise radii appropriate to the tracer density and threshold, and applied $R_{\rm eff}>R_{\rm shot}$ with $R_{\rm shot}\simeq 8\hmpc$ for $\Delta_{\rm lim}=-0.8$ and $R_{\rm shot}\simeq 9\hmpc$ for $\Delta_{\rm lim}=-0.7$ (see Section~\ref{sec:popcorn_data} and \citealt{paz_2022} for details). 
These cuts ensure that each catalog is volume-complete above the adopted scale and that small, noise-dominated detections are excluded.

To extend this threshold-based analysis to \zobov\ voids, we employ the tool \textsc{find\_radius\_and\_clean} provided within the VFT framework.
This method applies a nearest-neighbor based search around each void center to identify spherical regions that satisfy a predefined integrated density-contrast threshold and to assign an associated effective radius.
Voids for which such a region cannot be identified within a given search extent defined by the number of neighbors considered are discarded.
The choice of this extent is calibrated using the neighbor distributions characteristic of spherical void finders.
Finally, the remaining candidates are cleaned by removing overlapping regions, retaining only the largest non-overlapping voids with the biggest associated radius. 
The remaining catalog is not a pure spherical void but a \zobov\ cleaned catalog (tracers are conserved) with an associated radii such that the integrated density condition is fulfilled.
In this way, we construct spherical void samples derived from the \zobov\ catalog that satisfy integrated density thresholds of $\Delta_{\rm lim}=-0.9$, $-0.8$, and $-0.7$, allowing for a direct comparison with the threshold-based \sparkling\ and \popcorn\ voids.
For these derived \zobov\ samples, we apply the same selection criteria adopted for the \sparkling\ catalogs, retaining only voids with $R_{\rm void}>4\hmpc$, $4.5\hmpc$, and $5\hmpc$ for $\Delta_{\rm lim}=-0.9$, $-0.8$, and $-0.7$, respectively.
%

\begin{figure*}[h!]
\begin{center}
\includegraphics[width=\textwidth]{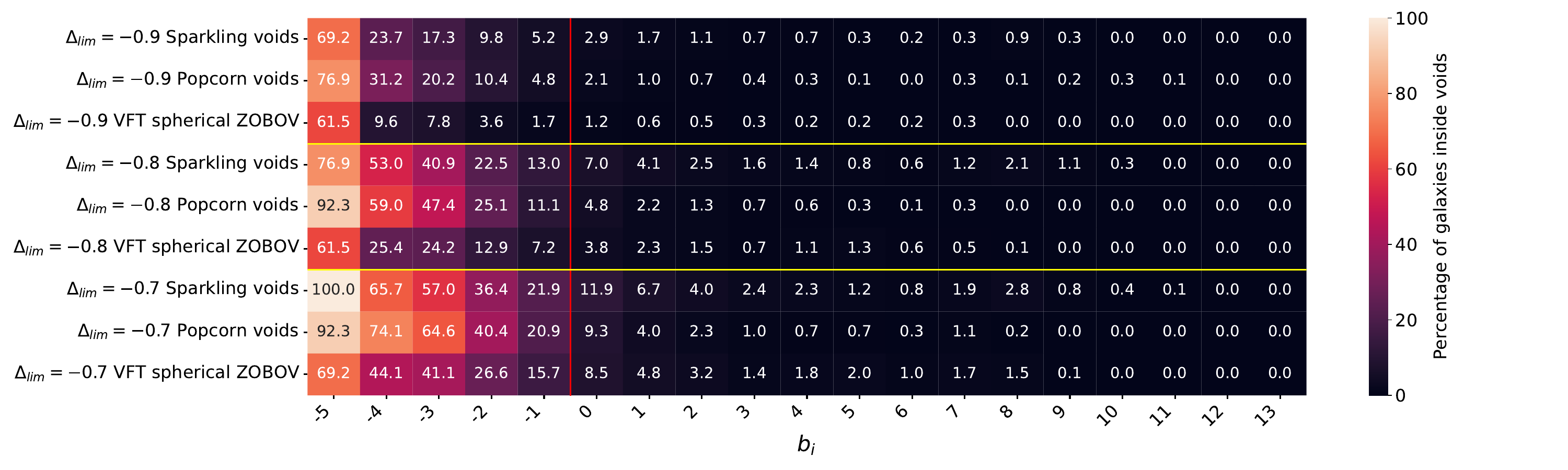}
\end{center}
\caption{\label{fig:bias_delta_matrix}
Correlation matrix between \sparkling\, \popcorn\ and VFT spherical \zobov\ voids identified with different density thresholds ($\Delta_{\rm lim} = -0.7, -0.8$ and $-0.9$). Each cell gives the fraction of galaxies in the corresponding $b_i$-bin that lie within each void catalog. The colors are given by the lateral bar. The red vertical line separates the negatives $b_i$-bins from the positives, and the horizontal yellow lines separate the void samples with different densities.}
\end{figure*}

%
For each resulting catalog (\sparkling, \popcorn, and VFT spherical \zobov\ voids at three density thresholds), we repeated the calculation of $b_i$–void occupancy introduced above: galaxies in the entire sample were binned by $b_i$, and for each $b_i$ bin we computed the fraction of galaxies that reside within the voids of each catalog.
The results are shown in Figure~\ref{fig:bias_delta_matrix}, where each row corresponds to a void catalog at a given $\Delta_{\rm lim}$ and each column to a $b_i$ bin; the colors and values of the cell indicate the percentage of galaxies in that $b_i$ bin that are contained within the catalog of voids specified.
Across all tested thresholds, the most negative $b_i$ bins exhibit the highest occupancy fractions for both \sparkling\ and \popcorn, confirming that integrated-density voids preferentially select anti-biased tracers.
Fully consistent behavior is observed for the VFT spherical \zobov\ voids, which display the same qualitative dependence on $b_i$ and on the adopted density threshold.
Moving from $\Delta_{\rm lim}=-0.9$ to $-0.8$ and $-0.7$ systematically increases the fraction of strongly negative-$b_i$ galaxies contained within the void catalogs.
This trend is present for all three definitions of voids and is more pronounced for \popcorn\ voids.
Thus, relaxing the underdensity criterion enhances the probability that strongly anti-biased galaxies are included in the void sample, independently of the specific threshold-based finder employed.
At every threshold, the \popcorn\ catalogs exhibit larger occupancy fractions in the most negative $b_i$ bins than the corresponding \sparkling\ and VFT spherical \zobov\ catalogs, while simultaneously showing reduced fractions of galaxies with $b_i \gtrsim 0$.
This behavior is consistent with the improved reconstruction of the void boundaries by \popcorn through the union of maximal spheres, which reduces contamination from the surrounding environment and increases the purity of the interior of the voids in terms of low-bias tracers.
The results in Figure~\ref{fig:bias_delta_matrix} show that density-threshold voids robustly isolate anti-biased galaxies.
Evermore, the link between individual galaxy anti-bias and these voids membership is robust and can be tuned by the void definition: less restrictive integrated underdensity thresholds yield larger void volumes that include a higher fraction of strongly anti-biased galaxies. 
To conclude, we note that the correlation observed here, relating the large-scale environment, characterized by $b_{i}$, to the density constraints of the void interior, could potentially be modeled analytically following the framework of \cite{shi2018}, which describes the conditional probability of the large-scale density field given a local small-scale underdensity constraint.
%

\section{Summary and conclusions}
\label{sec:summary}

In this work, we have conducted a systematic comparison of individual galaxy bias measurements, $b_i$, within cosmic voids identified by five distinct void-finding algorithms:
the spherical void finder \sparkling\ \citep{ruiz_void_2015,ruiz_into_2019}, the watershed-based \zobov\ algorithm \citep{neyrinck_2008}, the two modes of the \revolver\ framework (\revolvervoxel\ and \revolverzobov; \citealt{nadathur_2019}), and the free-form integrated-density approach \popcorn\ \citep{paz_2022}. 
By applying these complementary methodologies to the same underlying galaxy sample drawn from the IllustrisTNG simulation (TNG300-1 at $z=0$), we have assessed how the choice of void definition affects the observed environmental dependence of galaxy bias and the spatial distribution of anti-biased tracers.
Our analysis extends the results of \citet{montero-dorta_2025}, who demonstrated that galaxies residing in spherical voids exhibit systematically negative individual bias, with $\langle b_i \rangle$ increasing from the void center toward the boundary.
In that work, this radial dependence was identified as a characteristic void bias profile, capturing how galaxy--matter coupling varies across void interiors when distances are expressed in normalized void-centric units.
Here, we have shown that this fundamental trend is robust across multiple void identification schemes, although the amplitude and shape of the bias profile depend significantly on the adopted algorithm.
The main findings of this study can be summarized as follows:
\begin{itemize}
    \item 
    Density-threshold definitions (\sparkling\ and \popcorn) preferentially select galaxies with lower $b_i$ and produce distributions of $b_i$ shifted toward lower values relative to the full sample, whereas watershed-based methods tend to include a larger fraction of higher-bias, boundary galaxies.
    \item
    When radial distances are expressed in normalized units ($R/R_{\rm max}$), the majority of catalogs display a bias profile of voids that increases radially with comparable slopes, reinforcing the notion that the morphology of the profile, and not just its amplitude, is a reproducible property of void environments. \revolvervoxel\ is a notable exception that can be attributed to a geometric projection effect: because these voids are systematically smaller than those identified by other watershed methods, galaxies near their walls often lie deep within the interiors of larger \revolverzobov\ or \zobov\ structures, where bias values are still rising.
    \item
    Re-identifying voids at progressively weaker integrated-underdensity thresholds increases the fraction of low-bias galaxies contained within the resulting catalogs, with \popcorn\ providing the most effective isolation of low-bias tracers under the configurations tested.
\end{itemize}
These findings have important implications for observational and theoretical studies of cosmic voids and galaxy bias. 
First, they confirm that individual galaxy bias is a powerful diagnostic of galaxy–matter coupling that reveals environmental dependencies difficult to detect with traditional clustering statistics. 
Galaxies in the deepest underdensities exhibit systematically lower large-scale bias values, a signature of the distinct dynamical and evolutionary conditions prevailing in void interiors.
Second, our results demonstrate that the observed environmental modulation of bias is robust across void definitions. 
This robustness lends confidence to the physical reality of void-related bias trends and suggests that similar patterns should be detectable in observational data, despite survey-specific complications such as redshift-space distortions, photometric errors, and incompleteness.
Finally, the fact that normalizing by $R_{max}$ reveals similar void bias profiles across most finders suggests that the void size is a fundamental scale for characterizing the environmental dependence of bias, a feature that aligns naturally with excursion set peaks models \citep[e.g.][]{Sheth_VanDeWeygaert_2004}, where the formation of structures is tied to a characteristic smoothing scale, in contrast to scale-independent local bias expansions.
However, not all void catalogs are equally suited for bias studies.
Watershed-based methods that do not impose explicit density constraints, particularly standard \zobov, can generate void samples that include substantial high-density boundary material, leading to weaker or even inverted bias trends compared to density-threshold–based methods.
As a result, studies of galaxy properties in voids must carefully assess whether a given void definition is optimized for isolating deeply underdense environments or for capturing the full topological complexity of the cosmic web, as these objectives naturally select different galaxy populations.
In the broader context of galaxy evolution in low-density environments, numerous observational and theoretical works have shown that void galaxies tend to be bluer, more actively star-forming, and to host younger and more metal-poor stellar populations than their counterparts in average-density regions \citep[e.g.][]{Rojas2004, Rojas2005, vonBenda-Beckmann2008, Martizzi2020, Dominguez-Gomez2023a, Dominguez-Gomez2023b, Ceccarelli2022, Rosas_Guevara_2022, Curtis2024, Rodriguez-Medrano2023, Rodriguez-Medrano2024}. 
In particular, \citet{Rodriguez-Medrano2023} showed that, at fixed halo mass, void galaxies are systematically bluer, with higher SFRs and lower concentrations than their field counterparts, and that R/S void classification correlates with star-formation activity.
\citet{montero-dorta_2025} further reported differences in the individual-galaxy bias profile between R- and S-type voids.
More recently, MaNGA IFS results further corroborate younger, less metal-rich stellar populations in void galaxies \citep{Rodriguez-Medrano2025}.
These galaxy-level trends are reflected by halo-occupation diagnostics: halos above $\sim10^{12}\,h^{-1}M_\odot$ show a suppressed satellite occupation in void environments \citep{Alfaro2020,alfaro_hod_2022}, and recent work indicates a differential growth history for massive void halos consistent with reduced satellite accretion \citep{Alfaro2025}.
Together, these lines of evidence provide a coherent physical picture in which the dynamical state of the void environment (e.g., R/S classification) imprints signatures on both internal galaxy properties and large-scale bias, and our object-level measurements of $b_i$ offer a complementary manifestation of that process.
At the same time, the literature contains apparently discordant results regarding the magnitude of environmental effects \citep[e.g.][]{Kauffmann2004, Balogh2004, Dominguez2022}. 
Our findings help to clarify why the magnitude of these environmental effects has remained a matter of debate. 
By comparing void catalogs built with and without integrated-density constraints, we show that only voids associated with robust underdensities yield samples strongly dominated by anti-biased galaxies, whereas watershed-based catalogs tend to mix galaxies from a broader range of density environments. 
This methodological sensitivity naturally contributes to the diversity of conclusions in the literature: studies relying on void definitions that do not isolate deep underdensities are more likely to detect weaker or ambiguous environmental trends. 
In this sense, our bias measurements provide a unifying perspective, indicating that part of the apparent inconsistency regarding the influence of voids on galaxy evolution arises from differences in void-finder definitions rather than from the absence of a
physical effect.
Beyond these immediate results, the fact that a common, radially increasing voids bias profile with comparable slopes across the majority of void-finding methods is recovered once distances are expressed in normalized void-centric units supports its use as a robust ingredient in void-based cosmological analyses, where an accurate characterization of galaxy bias is essential. 
The ability to estimate large-scale galaxy bias on an object-by-object basis, and to characterize its behavior within voids, opens new avenues for extracting cosmological information related to the growth of structure \citep{Hamaus2016, Correa2019, Correa2021, Correa2022, Song2024, Contarini2024}.
In particular, a robust characterization of galaxy bias in and around voids may improve the modeling of void velocity profiles within the Gaussian streaming framework \citep{Fisher1995, Desjacques2010, paz_clues_2013, Hamaus2015}, as well as cosmological tests based on the abundance and properties of voids identified using biased tracers \citep{Contarini2019, paz_2022}.
Future work should extend this analysis in several directions. 
Exploring how individual galaxy bias within voids evolves with redshift and depends on galaxy properties (stellar mass, star formation rate, morphology) will provide crucial constraints on galaxy formation models in low-density regimes. 
In addition, developing hybrid void-finding approaches that combine the topological fidelity of watershed methods with the physical interpretability of density thresholds represents a promising avenue for constructing optimal void catalogs for large-scale structure studies.
In summary, we have demonstrated that the individual galaxy bias framework, when combined with a comprehensive comparison of void identification methods, offers a powerful tool for probing the interplay between cosmic environment and galaxy formation. 
Our results establish that galaxies with negative individual bias are preferentially located in cosmic voids, that this association is strongest for void definitions enforcing strict integrated underdensity criteria, and that the radial gradient of bias within voids is a robust feature across methodologies.
These findings deepen our understanding of how galaxies trace the matter distribution in the most extreme low-density regions of the Universe and provide a foundation for future observational and theoretical investigations of void galaxies and their role in cosmological tests.

\begin{acknowledgements}

This work has been partially supported with grants from Agencia Nacional de Promoción Científica y Tecnológica, the Consejo Nacional de Investigaciones Científicas y T\'ecnicas (CONICET, Argentina) and the Secretar\'{\i}a de Ciencia y Tecnolog\'{\i}a de la Universidad Nacional de C\'ordoba (SeCyT-UNC, Argentina). ADMD acknowledges support from the Universidad Técnica Federico Santa María through the Proyecto Interno Regular \texttt{PI\_LIR\_25\_04}. ADMD and FR acknowledge support from the Abdus Salam International Centre for Theoretical Physics through the Regular Associates Programme 2022–2027 and Junior Associates Programme 2023-2028, respectively. ABA acknowledges the Servicio Público de Empleo del Gobierno de España.
The IllustrisTNG simulations were undertaken with compute time awarded by the Gauss Centre for Supercomputing (GCS) under GCS Large-Scale Projects GCS-ILLU and GCS-DWAR on the GCS share of the supercomputer Hazel Hen at the High Performance Computing Center Stuttgart (HLRS), as well as on the machines of the Max Planck Computing and Data Facility (MPCDF) in Garching, Germany.
Figures were developed using \textsc{Matplotlib} \citep{Hunter2007} and some of them were post-processed with \textsc{Inkscape} (\url{https://inkscape.org}). 
\end{acknowledgements}

\bibliographystyle{aa}
\bibliography{references}

\end{document}